\documentclass{ws-ijmpd}
\IfFileExists{srcltx.sty}{\usepackage[active]{srcltx}}%

\begin{document}

%

\def\nocropmarks{\vskip5pt\phantom{cropmarks}}

 \let\trimmarks\nocropmarks      

%


\def\gtap{\mathrel{ \rlap{\raise 0.511ex \hbox{$>$}}{\lower 0.511ex
   \hbox{$\sim$}}}} \def\ltap{\mathrel{ \rlap{\raise 0.511ex
   \hbox{$<$}}{\lower 0.511ex \hbox{$\sim$}}}} 
\newcommand{\beq}{\begin{equation}}
\newcommand{\dd}{\partial}
\newcommand{\eeq}{\end{equation}}
\newcommand{\bea}{\begin{eqnarray}}
\newcommand{\eea}{\end{eqnarray}}
\newcommand{\lsim}{\stackrel{<}{\scriptstyle \sim}}
\newcommand{\gsim}{\stackrel{>}{\scriptstyle \sim}}
\newcommand{\La}{{\cal L} }
\newcommand{\half}{\frac{1}{2} }
\newcommand{\bra}[1]{\langle #1 |}
\newcommand{\ket}[1]{| #1 \rangle}
\newcommand{\sprod}[2]{\langle #1 , #2 \rangle}
\newcommand{\braket}[2]{\langle #1 | #2 \rangle}
\newcommand{\eq}[1]{eq.(\ref{#1})}
\newcommand{\dpar}[2]{\frac{\partial #1}{\partial #2}}
\newcommand{\vpar}[2]{\frac{\delta #1}{\delta #2}}
\newcommand{\ddpar}[2]{\frac{\partial^2 #1}{\partial #2^2}}
\newcommand{\vvpar}[2]{\frac{\delta^2 #1}{\delta #2^2}}
\newcommand{\eV}{\mbox{$ \ \mathrm{eV}$}}
\newcommand{\KeV}{\mbox{$ \ \mathrm{KeV}$}}
\newcommand{\MeV}{\mbox{$ \ \mathrm{MeV}$}}
\newcommand{\probm}{\mbox{$ \ \langle P_m \rangle$}}

\def\s{\mbox{\boldmath$\displaystyle\mathbf{\sigma}$}}
\def\J{\mbox{\boldmath$\displaystyle\mathbf{J}$}}
\def\K{\mbox{\boldmath$\displaystyle\mathbf{K}$}}

\def\beq{\begin{eqnarray}}
\def\eeq{\end{eqnarray}}

\def\cdash{$^{\raisebox{-0.5pt}{\hbox{--}}}$}  

%


\markboth{Alexander Kusenko}{Detecting Dark Matter in Space
}

%
\catchline{}{}{}
%

\title{Detecting sterile dark matter in space} 


\author{\footnotesize Alexander Kusenko} 

\address{Department of Physics and Astronomy, \\ 
         University of California, Los
          Angeles, CA 90095-1547, USA} 


\maketitle

\begin{abstract}

Space-based instruments provide new and, in some cases,
unique opportunities to search for dark matter.  In particular, if dark matter 
comprises sterile neutrinos, the x ray detection of their decay line is 
the most promising strategy for discovery.  Sterile neutrinos with
masses in the keV range could solve several long-standing astrophysical
puzzles, from supernova asymmetries and the pulsar kicks to star
formation, reionization, and baryogenesis. The best current limits on
sterile neutrinos come from Chandra and XMM-Newton.  Future advances can be
achieved with a high-resolution x-ray spectrometry in space.

\end{abstract}


\section{Introduction}

There is an overwhelming amount of evidence that most of the matter in the
universe is not made of ordinary atoms, but, rather, of new, yet undiscovered
particles\cite{dark_review}.  The evidence for dark matter is based on several
independent observations, including cosmic microwave background radiation,
gravitational lensing, the galactic rotation curves, and the x-ray observations
of clusters. None of the Standard Model particles can be dark matter.  Hence,
the identification of dark matter will be a discovery of new physics beyond the
Standard Model. 

To detect dark matter one must guess its properties, which ultimately
determine one's strategy for detection.   One can base one's guesses on
compelling theoretical ideas or on some observational clues.   

One of the most popular theories for physics beyond the Standard Model is
supersymmetry.  A class of supersymmetric extensions of the Standard Model
predict dark matter in the form of either the lightest supersymmetric
particles\cite{SUSY_dark_review}, or SUSY Q-balls\cite{Q_balls}.
Another theoretically appealing possibility is dark 
matter in the form of axions\cite{axion}.  Axion is a very weakly interacting
field which accompanies the Peccei--Quinn solution of the strong CP problem. 
There are several other dark-matter candidates that are well motivated by
theoretical reasoning.  A comprehensive review of possibilities is not our
purpose; rather, we will focus on the forms of dark matter which are 
well-motivated and for which there are new opportunities in space research.   

Right-handed or sterile neutrinos can be the dark
matter\cite{dw}\cdash\cite{nuMSM}. The existence of such right-handed states is
implied by the discovery of the active neutrino masses. Although it is not
impossible to explain the neutrino masses otherwise, most models introduce
gauge singlet fermions that give the neutrinos their masses via mixing.  If one
of these right-handed states has mass in the $\sim 1-50$~keV 
range, it can be the dark matter.  Several indirect astrophysical clues
support this hypothesis. Indeed, if the sterile neutrinos exist, they can
explain the long-standing puzzle of pulsar velocities\cite{Kusenko:review}.  In
addition, the x rays produced in decays of the relic neutrinos could increase
the ionization of the primordial gas and can catalyze the formation of
molecular hydrogen at redshift as high as 100. Since the molecular hydrogen is
an important cooling agent, its increased abundance could cause the early and
prompt start formation\cite{reion,reion1}. Sterile neutrinos can also help the
formation of supermassive black holes in the early universe\cite{mun}. For 
smaller masses, the sterile neutrinos have a long enough free-streaming
length to rectify several reported inconsistencies between the predictions of
cold dark matter on small scales and the observations.  The consensus of these
indirect observational hints helps make a stronger case for the sterile dark
matter.

\section{Sterile neutrinos}

The number of light ``active'' left-handed neutrinos -- three -- is well
established from the LEP measurements of the Z-boson decay width.  In the
Standard Model, the three active neutrinos fit into the three generations
of fermions. In its original form the Standard Model described massless
neutrinos.  The relatively recent but long-anticipated discovery of the
neutrino masses has made a strong case for considering right-handed
neutrinos, which are SU(3)$\times$SU(2)$\times$U(1) singlets.  The number
of right-handed neutrinos may vary and need not equal to
three.\cite{how_many} Depending on the structure of the
neutrino mass matrix, one can end up with none, one, or several states that
are light and (mostly) sterile, {\em i.e.}, they interact only through
their small mixing with the active neutrinos.

Sterile neutrino is not a new idea.   The name "sterile" was coined by
Bruno~Pontecorvo in 1967\cite{pontecorvo}.  Many seesaw models\cite{seesaw} 
assume that sterile neutrinos have very large masses, which makes them
unobservable.  However, one can consider a lighter sterile neutrino,
which can be dark matter\cite{dw}. Emission of sterile neutrinos
from a supernova could explain the pulsar kicks if the sterile neutrino mass
was several keV\cite{ks97,fkmp}.   More recently, a number of papers have
focused on this range of masses because several indirect observational hints
suggest the existence of a sterile neutrino with such a mass. 

Unless some neutrino experiments are wrong, the present data on neutrino
oscillations cannot be explained with only the active neutrinos.  Neutrino
oscillations experiments measure the differences between the squares of
neutrino masses, and the results are: one mass squared
difference is of the order of $10^{-5}$(eV$^2$), the other one is
$10^{-3}$(eV$^2$), and the third is about $1\,$(eV$^2$).  Obviously, one
needs more than three masses to get the three different mass splittings
which do not add up to zero.  Since we know that there are only three
active neutrinos, the fourth neutrino must be sterile.  However, if the
light sterile neutrinos exist, there is no compelling reason why their
number should be limited to one.

The neutrino masses can be introduced into the Standard Model by means of the
following addition to the lagrangian: 
\beq
{\cal L}
  = {\cal L_{\rm SM}}+\bar \nu_{s,a} \left(i \partial_\mu \gamma^\mu \right )
\nu_{s,a}
  - y_{\alpha a} H \,  \bar L_\alpha \nu_{s,a} 
  - \frac{M_{aa}}{2} \; \bar {\nu}_{s,a}^c \nu_{s,a} + h.c. \,,
\label{lagrangian}
\eeq
where $H$ is the Higgs boson and $L_\alpha$ ($\alpha=e,\mu,\tau$) are the
lepton doublets, while $ \nu_{s,a}$ ($a=1,...,N$) are the additional singlets.
This model, dubbed $\nu$MSM\cite{nuMSM}, provides a natural framework for
considering sterile neutrinos.  Of course, the gauge singlet fields may have
some additional couplings omitted from eq.~(\ref{lagrangian}).  
 The neutrino mass matrix has the form 
\beq
M=\left (
\begin{array}{cc}
\tilde{m}_{3\times 3} & D_{3\times N} \\
D_{N \times 3}^T & M_{ N \times N} 
\end{array}
\right ), 
\label{massmatrix}
\eeq
where the Dirac masses $D_{\alpha a}= y_{\alpha a} \langle H \rangle $ are the
result of spontaneous symmetry breaking.   For symmetry reasons one usually sets
$\tilde{m}_{3\times 3}$ to zero.   As for the right-handed Majorana masses $M$,
the scale of these masses can be either much greater or much smaller than the
electroweak scale.   

The seesaw mechanism\cite{seesaw} can explain the smallness of neutrino masses
in the presence of the Yukawa couplings of order one.  For this purpose, one
assumes that the Majorana masses are much larger than the electroweak scale,
and the smaller eigenvalues of the mass matrix~(\ref{massmatrix}) are suppressed
by the ratio of $ \langle H \rangle $ to $M$.  

However, the origin of the Yukawa couplings remains unknown, and, in the
absence of the fundamental theory, there is no compelling reason to believe
that these couplings must be of order 1.  Indeed, the Yukawa couplings of most
known fermions are much smaller than one, e.g. the Yukawa coupling of the
electron is $\sim 10^{-6}$. 

Thus, for all we know, the scale of the Majorana mass $M$ in
eq.~(\ref{massmatrix}) can be much smaller than th electroweak scale.  If
$M\sim 1$~eV, the sterile
neutrinos with the mass $m_s\sim 1$~eV can explain the LSND
results\cite{deGouvea:2005er}.  If $M\sim 1$~keV,  the sterile neutrinos with
the corresponding mass could explain the pulsar kicks\cite{ks97,fkmp} and dark
matter\cite{dw}, and they can also play a role in generating the
matter-antimatter asymmetry of the universe\cite{baryogenesis}.  

\section{Production of sterile neutrinos in the early universe}

Sterile neutrinos can be produced in the early universe from neutrino
oscillations, as well as from other couplings, not included in
eq.~(\ref{lagrangian}).   For example, dark matter in the form of sterile
neutrinos can be produced by a direct coupling to the
inflaton\cite{Shaposhnikov:2006xi}.

At very high temperatures the active neutrinos have frequent interactions in
plasma, which reduce the probability of their conversions into sterile
neutrinos\cite{stodolsky}. The mixing of sterile neutrinos with one of the
active species in plasma 
can be represented by an effective, density and temperature dependent 
mixing angle\cite{dw}\cdash\cite{dolgov_hansen}:
\begin{eqnarray}
| \nu_1 \rangle & = & \cos \theta_m \, | \nu_e \rangle - \sin \theta_m \, |
\nu_s  \rangle \\ 
| \nu_2 \rangle & = & \sin \theta_m \, | \nu_e \rangle + \cos \theta_m \, |
\nu_s \rangle  ,
\label{eigenstates}
\end{eqnarray}
where
\begin{equation}
\sin^2 2 \theta_m = 
\frac{(\Delta m^2 / 2p)^2 \sin^2 2 \theta}{(\Delta m^2 / 2p)^2 \sin^2 
2 \theta + ( \Delta m^2 / 2p \cos 2 \theta - V_m-V_{_T})^2}. 
\label{sin2theta}
\end{equation}
Here $V_m$ and $V_T$ are the effective matter and temperature potentials.
In the limit of small angles and small lepton asymmetry, the mixing angle
can be approximated as 
\beq
\sin 2 \theta_m \approx
\frac{\sin 2 \theta}{1+ 0.27 \zeta  \left( \frac{T}{100 \,
\rm MeV} \right)^6 \left( \frac{{\rm keV}^2}{\Delta m^2} \right )
}
\eeq
where $\zeta =1.0$ for mixing with the electron neutrino and  $\zeta 
=0.30$ for $\nu_\mu$ and $\nu_\tau$.  

Obviously, thermal effects suppress the mixing significantly for
temperatures $T> 150 \, (m/{\rm keV})^{1/3}\,$MeV.  If the singlet
neutrinos interact only through mixing, all the interaction rates are
suppressed by the square of the mixing angle, $\sin^2 \theta_m $.  It is
easy to see that these sterile neutrinos are {\em \sf never} in thermal
equilibrium in the early universe.  Thus, in contrast with the case of the
active neutrinos, the relic population of sterile neutrinos is not a result
of a freeze-out.  One immediate consequence of this observation is that the
Gershtein--Zeldovich bound\cite{gz} and the Lee--Weinberg bound\cite{lw} do
not apply to sterile neutrinos.  In general, the existing experimental
constraints on sterile neutrinos\cite{sterile_constraints} allow a wide range
of parameters, especially for small mixing angles. 

\begin{figure}[ht]
\centerline{\epsfxsize=4 in\epsfbox{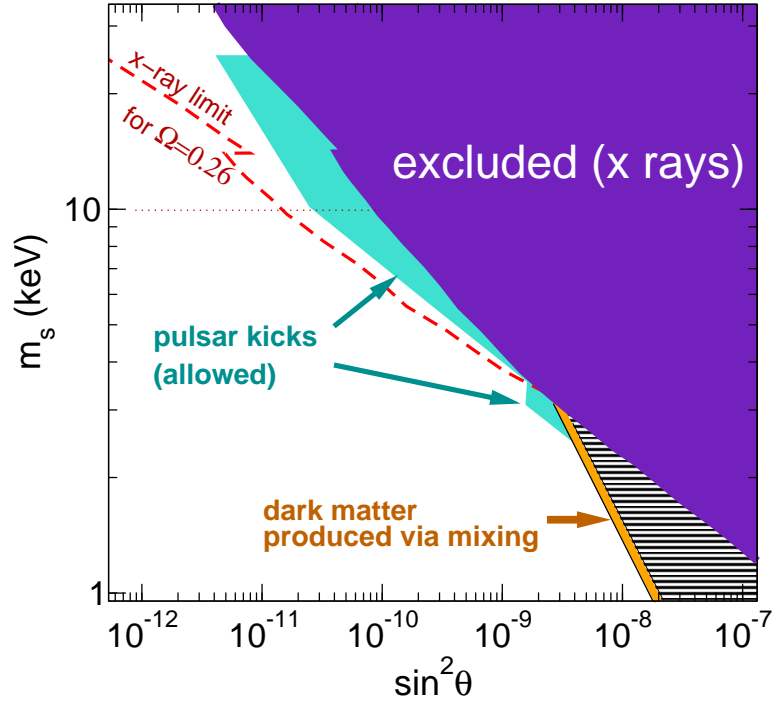}}   
\caption{The range of the sterile neutrino masses and mixing angles.  The x-ray
limits depend on the abundance of the relic sterile neutrinos, which, in turn,
depends on their production mechanism.  If the sterile neutrinos are produced
only via their mixing with active neutrinos, they can be dark matter for
masses below 3~keV, as shown in the figure.  This range is in conflict with
the 10~keV lower bound from the Lyman-alpha forest shown by a dotted line. 
In sharp contrast, the observations of dwarf spheroid galaxies favor the
masses of a few keV.  If the sterile neutrinos are produced via some additional
couplings, besides the mixing with the active neutrinos, and if the sterile
neutrinos make up all the dark matter ($\Omega=0.26$), the corresponding x-ray
limit is shown as a dashed line. Also shown is the allowed range of parameters
consistent with the pulsar kicks. }
\label{figure:range}
\end{figure}

 One can calculate the production of
sterile neutrinos in plasma by solving the Boltzmann equation for the
distribution function $f(p,t)$:
\bea
\left ( \frac{\dd}{\dd t}-H p \frac{\dd}{\dd p}\right ) f_s(p,t) & \equiv
& x H\dd_x f_s =  \\
& & \Gamma_{(\nu_a\rightarrow \nu_s)}
\left ( f_a(p,t) - f_s(p,t) \right ),  
\eea
where $H$ is the Hubble constant, $x=1\,$MeV~$a(t)$, $a(t)$ is the scale
\cite{Abazajian:2005gj}
factor, and $\Gamma$ is the probability of conversion.  
The solution\cite{dw}\cdash\cite{dolgov_hansen,Abazajian:2005gj}  is
shown in Fig.\ref{figure:range} as "dark matter produced via mixing".  
One should keep in mind that this solution is subject to hadronic 
uncertainties\cite{Shaposhnikov:2006rw}. 

If the sterile neutrinos have additional interactions, not included in
eq.~(\ref{lagrangian}), the relic population of these particles can be
produced via different mechanisms.  One example is a direct coupling of
sterile neutrinos to the inflaton\cite{Shaposhnikov:2006xi}.  In this case the
production of sterile neutrinos may not be governed by the mixing angle,
although the mixing angle still controls the decay rate and, therefore, some
of the constraints discussed below depend on the mixing angles. 

\section{Constraints on sterile dark matter} 

Although dark-matter sterile neutrinos are stable on cosmological time scales,
they nevertheless decay\cite{dolgov_hansen,x-rays}.  The dominant decay mode,
into three light neutrinos,
is "invisible" because the daughter neutrinos are beyond the detection
capabilities of today's experiments.  The most prominent "visible" mode is
decay into one active neutrino and one photon, $\nu_s \rightarrow \nu_a
\gamma$.  Assuming the two-neutrino mixing for simplicity, one can express the
inverse width of such a decay as\cite{pal_wolf}
\begin{equation}
\tau \equiv \Gamma_{\nu_s \rightarrow \nu_a
\gamma}^{-1}= 1 \times 10^{26} \rm{ s } \left(\frac{7\ \rm keV}{m_{s}
}\right)^{5} \left(\frac{1\times 10^{-9}}{\sin^{2}
\theta}\right),  
\label{meanlife}
\end{equation}
where $m_s$ is the mass and $\theta $ is the mixing angle.  

Since this is a two-body decay, the photon energy is half the mass of
the sterile neutrino.  The monochromatic line from dark matter decays can, in
principle, be observed by x-ray telescopes.  No such observation has been
reported, and some important limits have been derived on the allowed masses and
mixing angles.   These constraints
are based on different astrophysical objects, from Virgo and Coma clusters, to
Large Magellanic Clouds, to Milky Way halo and its components\cite{x-rays}. 
There are
different uncertainties in modeling the dark matter populations in these
objects. Different groups have also used very different methods in
deriving these bounds: from a conservative assumption that the dark-matter line
should not exceed the signal, to more ambitious approaches that involved
modeling the signal or merely fitting it with a smooth curve and requiring that
the line-shaped addition not affect the quality of the fit.  In any case, the
limits apply to the flux of x rays, which can be translated into the limits on
mass and mixing angle if the sterile neutrino abundance is known.  As we
discussed above, the production is possible via the mixing alone, but some
additional couplings and other production mechanisms are by no means
excluded\cite{Shaposhnikov:2006xi}. 
Most published bounds\cite{x-rays} {\em \sf assume} that sterile neutrinos make
up
all the dark matter, that is $ \Omega_s = 0.26$.  The limit based on this
assumption is shown as a dashed (red) line in Fig.~1.  However, it should not
be used as the exclusion limit for sterile neutrinos in general, because it is
possible that  $ \Omega_s <  0.26$, while the sterile neutrinos could still
explain the pulsar velocities and they could play a role in the star
formation.  

A different kind of limit based on the same x-ray data\cite{x-rays} can be
set without assuming $ \Omega_s = 0.26$.   As long as there is a mixing due to
the couplings in the lagrangian~(\ref{lagrangian}), some sterile neutrinos
are produced in the hot plasma, regardless of any additional couplings that
may or may not be present. This amount corresponds to the lower bound on the
sterile neutrino abundance, and the bound obtained this way is the most robust,
model-independent limit. The corresponding exclusion region is shown in Fig.~1.

Additional constraints on dark matter come from the observations of the
Lyman-alpha forest\cite{viel1}\cdash\cite{viel2}, which limit the sterile
neutrino mass from below. Based on the high-redshift data from SDSS and some 
modeling of gas dynamics, one can set a limit as strong as
14~keV\cite{seljak}. However, the high-redshift data may have systematic
errors, and more conservative approaches, 
based on the relatively low-redshift data, have led to some less stringent
bounds\cite{viel1}.  Recently Viel et al.\cite{viel2} have reanalyzed the
high-redshift data and arrived at the bound $m_s>10$~keV.  The mass bounds as
quoted depend on the production mechanism in the early universe.  

The Lyman-alpha observations constrain the free-streaming lengths of dark
matter particles, not their masses.  For each cosmological production
mechanism, the relation between the free-streaming length and the mass is
different\cite{Asaka:2006ek}.  For example, the bound $m_s >10$~keV\cite{viel2}
applies to the production model due to Dodelson and Widrow\cite{dw}. If the
lepton asymmetry of the universe (which is unknown {\em a priori}) is
sufficiently
large, then the sterile neutrinos can be produced through resonant
Mikheev-Smirnov-Wolfenstein\cite{msw} (MSW)
oscillations in the early universe\cite{shi_fuller}. These neutrinos are
non-thermal and colder because the adiabaticity condition selects the
low-energy
part of the neutrino spectrum. Even within a given cosmological 
scenario, there are uncertainties in the production rates of neutrinos for any
given mass and mixing angle\cite{Shaposhnikov:2006rw}.  These uncertainties may
further affect the interpretation of the Lyman-alpha bounds in terms of the
sterile neutrino mass. 

It should also be mentioned that the Lyman-alpha bounds appear to contradict
the observations of dwarf spheroidal
galaxies\cite{Strigari:2006ue},
which  suggest that dark matter is warm and which would favor the 1--5~keV mass
range for sterile neutrinos. There are several inconsistencies between the
predictions of N-body simulations of cold dark matter (CDM) and the
observations\cite{cdm-wdm}.
Each of these problems may find a separate independent solution.   Perhaps, a
better understanding of CDM on small scales will resolve these discrepancies.  
It is true, however, that warm dark matter in the form of sterile neutrinos
is free from all these small-scale problems altogether, while on large scales
WDM fits the data as well as CDM.

If the sterile neutrinos make up only a part of dark matter, 
 the Lyman-alpha bounds do not apply.  In this case, the sterile neutrinos may
still be responsible for pulsar velocities, and they can play a role in  star
formation and reionization of the universe.  Also, if inflation ended with a
low reheat temperature, the bounds are significantly weaker\cite{low-reheat}.

\section{Reionization and star formation}

Sterile neutrinos decay in the early universe, in particular, during
the "dark ages" following recombination. The ionizing photons are too few to
affect the cosmic microwave background directly\cite{mapelli}, but they can
have an important effect on star formation and reionization.   The star
formation requires
cooling and collapse of gas clouds, which is impossible unless the fraction of
molecular hydrogen is high enough\cite{tegmark97}.  Star formation is
accompanied by the reionization of gas in the universe.  The WMAP (three years)
measurement\cite{Spergel:2006} of the reionization redshift
$z_r=10.9^{+2.7}_{-2.3}$ has posed a new challenge to theories of star
formation. On the one hand, stars have to form early enough to reionize gas at
redshift~11. On the other hand, the spectra of bright distant quasars imply
that reionization must be completed by redshift~6. Stars form in clouds of
hydrogen, which collapse at different times, depending on their sizes: the
small clouds collapse first, while the large ones collapse last. If the big
clouds must collapse by redshift~6, then the small halos must  undergo the
collapse at an earlier time.  It appears that the star formation in these 
small halos would have occurred at high redshift, when the gas density was very
high, and it would have resulted in an unacceptable overproduction of the
Thompson optical depth \cite{haiman_bryan}. To be consistent with WMAP, the
efficiency for the production of ionizing photons in minihalos must have been
at
least an order of magnitude lower than expected\cite{haiman_bryan}. One
solution is to suppress the star formation rate in small halos by some
dynamical
feedback mechanism. The suppression required is by at least an order of
magnitude.

An alternative solution is to consider warm dark matter, in which case the
small clouds are absent altogether. However, it has been argued that "generic"
warm dark matter can delay the collapse of gas clouds\cite{yoshida}.  This
problem does not arise in the case of
sterile neutrinos, because the x-ray photons from their slow decays could have 
increased the production of molecular hydrogen and could have precipitated a
rapid  and prompt star formation at a high enough redshift\cite{reion,reion1}.

\section{Pulsar velocities}

The space velocities of pulsars 
range from 250~km/s to 500~km/s.\cite{astro1}\cdash\cite{astro_15} 
Some 15\% of pulsars\cite{astro_15} appear to have velocities greater than
1000~km/s, while the fastest pulsars have speeds as high as 1600~km/s. The
origin of these velocities remains a puzzle\cite{Kusenko:review}.  
Since most of the supernova energy, as much as 99\% of the total $10^{53}$~erg
are emitted in neutrinos, a few per cent anisotropy in the distribution of
these neutrinos would be sufficient to explain the pulsar kicks.

Neutrinos are always {\em \sf produced } with an asymmetry, but they
usually {\em \sf escape} isotropically.  The asymmetry in production comes
from the asymmetry in the basic weak interactions in the presence of a
strong magnetic field\footnote{Here we disregard the neutrino magnetic
moments, which are negligible in the Standard Model and its simplest
extensions. Even for vanishing magnetic moments, neutrino oscillations are
affected by the magnetic field through the polarization of the matter
fermions\cite{osc_matter_B}}. Indeed, if the electrons and other fermions are
polarized by the magnetic field, the cross section of the urca processes,
such as $n+e^+ \rightleftharpoons p+ \bar \nu_e$ and
$p+e^-\rightleftharpoons n+ \nu_e $, depends on the orientation of the
neutrino momentum:
\beq 
\sigma ( \uparrow e^-, \uparrow \nu ) \neq  \sigma ( \uparrow e^-,
\downarrow \nu ) 
\label{sigma_up_down}
\eeq
Depending on the fraction of the electrons in the lowest Landau level, this
asymmetry can be as large as 30\%, which is, seemingly, more than one needs
to explain the pulsar kicks.\cite{drt}  However, this asymmetry is
completely washed out by scattering of neutrinos on their way out of the
star.\cite{eq}  This is intuitively clear because, as a result of
scatterings, the neutrino momentum is transferred to and shared by the
neutrons.  In the approximate thermal equilibrium, no asymmetry in the
production or scattering amplitudes can result in a macroscopic
momentum anisotropy.  This statement can be proved rigorously\cite{eq}.  

However, if the neutron star cooling produced a particle whose interactions
with nuclear matter were {\em \sf even weaker} than those of ordinary
neutrinos, such a particle could escape the star with an anisotropy equal
its production anisotropy.  The sterile neutrinos, whose interactions
are suppressed by $(\sin^2 \theta_m)$ can play such a
role\cite{ks97,fkmp,barkovich}. 
The region of masses and mixing angles consistent with this explanation for
the pulsar kicks is shown in Fig.~1.  The neutrino-driven kicks have a number
of ramifications: in particular, they can increase the energy of the shock and
can generate asymmetric jets, the strongest of which is aligned with the
direction of the pulsar motion\cite{Fryer}.

\section{Conclusion} 

Several independent observational hints point to sterile neutrinos with masses
in the keV range.  Pulsar velocities can be explained by the emission of such
sterile neutrinos from a supernova, because the sterile neutrino emission is
anisotropic in the presence of the magnetic field. The x-ray photons from the
decays of the sterile neutrinos can ionize the primordial gas and can cause an
increase in the fraction of molecular hydrogen, which makes a prompt star
formation possible at a relatively high redshift.   The sterile neutrinos
can be the dark matter.  In addition, they could have played a role in
generating the matter-antimatter asymmetry of the universe. Future observations
of x-ray telescopes may be able to discover the relic
sterile neutrinos by detecting keV photons from their decays.

\section*{Acknowledgments}
This work was supported in part by the
U.S. Department of Energy grant DE-FG03-91ER40662 and by the NASA ATP grants
NAG~5-10842 and NAG~5-13399.



\begin{thebibliography}{99}

\bibitem{dark_review}
  For review, see, e.g., G.~Bertone, D.~Hooper and J.~Silk,
  Phys.\ Rept.\  {\bf 405}, 279 (2005).

\bibitem{SUSY_dark_review}
For review, see, e.g., G.~Jungman, M.~Kamionkowski and K.~Griest,
  Phys.\ Rept.\  {\bf 267}, 195 (1996).
  
  \bibitem{Q_balls}
    A.~Kusenko,
  Phys.\ Lett.\ B {\bf 405}, 108 (1997); 
    A.~Kusenko and M.~E.~Shaposhnikov,
  Phys.\ Lett.\ B {\bf 418}, 46 (1998); 
    A.~Kusenko, V.~Kuzmin, M.~E.~Shaposhnikov and P.~G.~Tinyakov,
  Phys.\ Rev.\ Lett.\  {\bf 80}, 3185 (1998). 
%
For review, see, e.g.,   K.~Enqvist and A.~Mazumdar,
  Phys.\ Rept.\  {\bf 380}, 99 (2003); 
  M.~Dine and A.~Kusenko,
  Rev.\ Mod.\ Phys.\  {\bf 76}, 1 (2004).

  \bibitem{axion}
  R.~D.~Peccei and H.~R.~Quinn,
  Phys.\ Rev.\ Lett.\  {\bf 38}, 1440 (1977); 
  Phys.\ Rev.\ D {\bf 16}, 1791 (1977); 
%
  S.~Weinberg,
  Phys.\ Rev.\ Lett.\  {\bf 40}, 223 (1978); 
  F.~Wilczek,
  Phys.\ Rev.\ Lett.\  {\bf 40}, 279 (1978).
  
  
\bibitem{dw} 
S.~Dodelson and L.~M.~Widrow, 
Phys.\ Rev.\ Lett.\  {\bf 72}, 17 (1994). 

\bibitem{Fuller}
K.~Abazajian, G.~M.~Fuller and M.~Patel,
Phys.\ Rev.\ D {\bf 64}, 023501 (2001)

\bibitem{dolgov_hansen} 
A.~D.~Dolgov and S.~H.~Hansen,
Astropart.\ Phys.\  {\bf 16}, 339 (2002). 

\bibitem{nuMSM}
  T.~Asaka, S.~Blanchet and M.~Shaposhnikov,
  Phys.\ Lett.\ B {\bf 631}, 151 (2005)

\bibitem{Kusenko:review}
  A.~Kusenko,
  Int.\ J.\ Mod.\ Phys.\ D {\bf 13}, 2065 (2004).
  
\bibitem{reion}
  P.~L.~Biermann and A.~Kusenko,
 Phys.\ Rev.\ Lett.\  {\bf 96}, 091301 (2006).

\bibitem{reion1}
  J.~Stasielak, P.~L.~Biermann and A.~Kusenko,
  arXiv:astro-ph/0606435.
  
\bibitem{baryogenesis}
  E.~K.~Akhmedov, V.~A.~Rubakov and A.~Y.~Smirnov,
  Phys.\ Rev.\ Lett.\  {\bf 81}, 1359 (1998);  
  T.~Asaka and M.~Shaposhnikov,
  Phys.\ Lett.\ B {\bf 620}, 17 (2005).

\bibitem{mun}
  F.~Munyaneza and P.~L.~Biermann,
  arXiv:astro-ph/0403511.
  
\bibitem{how_many}
P.~H.~Frampton, S.~L.~Glashow and T.~Yanagida,
Phys.\ Lett.\ B {\bf 548}, 119 (2002); 
B.~Kayser,
Nucl.\ Phys.\ Proc.\ Suppl.\  {\bf 118}, 425 (2003).
  
  
\bibitem{pontecorvo}
B.~Pontecorvo, JETP, 53, 1717 (1967). 

  
\bibitem{seesaw} P. Minkowski, Phys. lett. {\bf B67 }, 421
(1977); M.~Gell-Mann, P.~Ramond, and R.~Slansky, \emph{Supergravity}
(P.~van Nieuwenhuizen et al. eds.), North Holland, Amsterdam, 1980,
p.~315; T.~Yanagida, in \emph{Proceedings of the Workshop on the
Unified Theory and the Baryon Number in the Universe} (O.~Sawada
and A.~Sugamoto, eds.), KEK, Tsukuba, Japan, 1979, p.~95; S.~L.
Glashow, \emph{The future of elementary particle physics}, in
   \emph{Proceedings of the 1979 Carg{\`e}se Summer Institute on
Quarks and Leptons} (M.~L{\'e}vy et al. eds.), Plenum Press, New York,
1980, pp.~687; R.~N. Mohapatra and G.~Senjanovi{\'c}, Phys. Rev. Lett.
\textbf{44}, 912 (1980).


\bibitem{ks97}
  A.~Kusenko and G.~Segr\`e, 
Phys.\ Lett.\ B {\bf 396}, 197 (1997); 
A.~Kusenko and G.~Segre,
  Phys.\ Rev.\ D {\bf 59}, 061302 (1999).

\bibitem{fkmp}
G.~M.~Fuller, A.~Kusenko, I.~Mocioiu, and S.~Pascoli,
Phys.\ Rev.\ D {\bf 68}, 103002 (2003). 
  
\bibitem{deGouvea:2005er}
  A.~de Gouvea,
  Phys.\ Rev.\ D {\bf 72}, 033005 (2005).

\bibitem{Shaposhnikov:2006xi}
  M.~Shaposhnikov and I.~Tkachev,
  arXiv:hep-ph/0604236.


\bibitem{stodolsky}
L.~Stodolsky,
Phys.\ Rev.\ D {\bf 36}, 2273 (1987); 
  R.~Barbieri and A.~Dolgov,
  Nucl.\ Phys.\ B {\bf 349}, 743 (1991).
  
\bibitem{gz}
S.~S.~Gershtein and Y.~B.~Zeldovich,
JETP Lett.\  {\bf 4} (1966) 120
[Pisma Zh.\ Eksp.\ Teor.\ Fiz.\  {\bf 4} (1966) 174].

\bibitem{lw}
B.~W.~Lee and S.~Weinberg,
Phys.\ Rev.\ Lett.\  {\bf 39}, 165 (1977).

\bibitem{sterile_constraints}
  A.~Kusenko, S.~Pascoli and D.~Semikoz, 
  JHEP {\bf 0511}, 028 (2005).

\bibitem{Abazajian:2005gj}
  K.~Abazajian,
  Phys.\ Rev.\ D {\bf 73}, 063506 (2006).

\bibitem{Shaposhnikov:2006rw}
  T.~Asaka, M.~Laine and M.~Shaposhnikov,
  JHEP {\bf 0606}, 053 (2006). 
  
  \bibitem{x-rays}
K.~Abazajian, G.~M.~Fuller and W.~H.~Tucker,
Astrophys.\ J.\  {\bf 562}, 593 (2001); 
  A.~Boyarsky, A.~Neronov, O.~Ruchayskiy and M.~Shaposhnikov,
  arXiv:astro-ph/0512509; 
  A.~Boyarsky, A.~Neronov, O.~Ruchayskiy and M.~Shaposhnikov,
  JETP Lett.\  {\bf 83}, 133 (2006); 
A.~Boyarsky, A.~Neronov, A.~Neronov, O.~Ruchayskiy and M.~Shaposhnikov,
  arXiv:astro-ph/0603368; 
 A.~Boyarsky, A.~Neronov, O.~Ruchayskiy, M.~Shaposhnikov and I.~Tkachev,
  arXiv:astro-ph/0603660; 
S.~Riemer-Sorensen, S.~H.~Hansen and K.~Pedersen,
  arXiv:astro-ph/0603661; 
  K.~Abazajian and S.~M.~Koushiappas,
  arXiv:astro-ph/0605271; 
  C.~R.~Watson, J.~F.~Beacom, H.~Yuksel and T.~P.~Walker,
  arXiv:astro-ph/0605424.

\bibitem{pal_wolf}
  P.~B.~Pal and L.~Wolfenstein,
  Phys.\ Rev.\ D {\bf 25}, 766 (1982).
  
  \bibitem{viel1}
M.~Viel, J.~Lesgourgues, M.~G.~Haehnelt, S.~Matarrese and A.~Riotto,
  Phys.\ Rev.\ D {\bf 71}, 063534 (2005). 
    
\bibitem{seljak}
  U.~Seljak, A.~Makarov, P.~McDonald and H.~Trac,
  arXiv:astro-ph/0602430.
  
\bibitem{viel2}
  M.~Viel, J.~Lesgourgues, M.~G.~Haehnelt, S.~Matarrese and A.~Riotto,
  arXiv:astro-ph/0605706.
  
\bibitem{Asaka:2006ek}
  T.~Asaka, A.~Kusenko and M.~Shaposhnikov,
  Phys.\ Lett.\ B {\bf 638}, 401 (2006).
  
  \bibitem{msw} 
S. P.~Mikheev and A.~Yu.~Smirnov, Yad. Fiz. {\bf 42}, 1441 
(1985) [Sov. J. Nucl. Phys. {\bf 42}, 913 (1985)]; L.~Wolfenstein,
Phys. Rev. {\bf D 17}, 2369 (1978). 


\bibitem{shi_fuller}
X.~d.~Shi and G.~M.~Fuller,
Phys.\ Rev.\ Lett.\  {\bf 82}, 2832 (1999).

\bibitem{cdm-wdm}
  G.~Kauffmann, S.~D.~M.~White and B.~Guiderdoni,
  Mon.\ Not.\ Roy.\ Astron.\ Soc.\  {\bf 264}, 201 (1993); 
  A.~A.~Klypin, A.~V.~Kravtsov, O.~Valenzuela and F.~Prada,
  Astrophys.\ J.\  {\bf 522}, 82 (1999); 
  B.~Moore, S.~Ghigna, F.~Governato, G.~Lake, T.~Quinn, J.~Stadel and P.~Tozzi,
ApJ {\bf 524}, L19 (1999).   
  B.~Willman, F.~Governato, J.~Wadsley and T.~Quinn,
MNRAS, 355, 159 (2004); 
  P.~Bode, J.~P.~Ostriker and N.~Turok,
  Astrophys.\ J.\  {\bf 556}, 93 (2001). 
  P.~J.~E.~Peebles,
ApJ, {\bf 557}, 495 (2001); 
  J.~J.~Dalcanton and C.~J.~Hogan,
  Astrophys.\ J.\  {\bf 561}, 35 (2001);
  A.~R.~Zentner and J.~S.~Bullock,
  Phys.\ Rev.\ D {\bf 66}, 043003 (2002); 
  F.~Governato {\it et al.},
  Astrophys.\ J.\  {\bf 607}, 688 (2004); 
    G.~Gentile, P.~Salucci, U.~Klein, D.~Vergani and P.~Kalberla,
  Mon.\ Not.\ Roy.\ Astron.\ Soc.\  {\bf 351}, 903 (2004); 
  J.~Kormendy, M.~E.~Cornell, D.~L.~Block, J.~H.~Knapen and E.~L.~Allard,
  arXiv:astro-ph/0601393.
  

\bibitem{Strigari:2006ue}
  M.~I.~Wilkinson {\it et al.},
  arXiv:astro-ph/0602186; 
  L.~E.~Strigari, J.~S.~Bullock, M.~Kaplinghat, A.~V.~Kravtsov, O.~Y.~Gnedin,
K.~Abazajian and A.~A.~Klypin,
  arXiv:astro-ph/0603775.
  
  
\bibitem{low-reheat}
  G.~Gelmini, S.~Palomares-Ruiz and S.~Pascoli,
  Phys.\ Rev.\ Lett.\  {\bf 93}, 081302 (2004).
  
\bibitem{mapelli}
  M.~Mapelli, A.~Ferrara and E.~Pierpaoli,
  Mon.\ Not.\ Roy.\ Astron.\ Soc.\  {\bf 369}, 1719 (2006).
  
  \bibitem{tegmark97} 
  M.~Tegmark, J.~Silk, M.~J.~Rees, A.~Blanchard, T.~Abel and F.~Palla,
  Astrophys.\ J.\  {\bf 474}, 1 (1997).
 
\bibitem{Spergel:2006}
  D.~N.~Spergel {\it et al.},
  arXiv:astro-ph/0603449.
  
\bibitem{haiman_bryan}
  Z.~Haiman and G.~L.~Bryan,
  arXiv:astro-ph/0603541.

\bibitem{yoshida} N.~Yoshida, A. Sokasian, L.~Hernquist, and V.~Springel,
Astrophys. J. Lett., {\bf 591}, 1 (2003).
  
\bibitem{astro1} A.~G.~Lyne, B.~Anderson, and
M.~J.~Salter, Mon. Not. R. Astron. Soc. {\bf 201}, 503 (1982); Bailes {\em
et al.}, Astrophys. J. {\bf 343}, L53 (1989); Formalont {\em et al.},
Mon. Not. R. Astron. Soc. {\bf 258}, 497 (1992); P.~A.~Harrison,
A.~G.~Lyne, and B.~Anderson, Mon. Not. R. Astron. Soc. {\bf 261} 113
(1993); A.~G.~Lyne and D.~R.~Lorimer, Nature 369 (1994) 127; 
%
P.~A.~G.~Scheuer, Nature {\bf 218}, 920 (1968);
  B.~J.~Rickett,  Mon. Not. R. Astron. Soc. {\bf 150}, 67 (1970); 
%
J.~A.~Galt and A.~G.~Lyne,
  Mon. Not. R. Astron. Soc. {\bf 158}, 281 (1972); 
Slee {\em et al}, {\em ibid.} {\bf 167}, 31 (1974); A.~G.~Lyne and
  F.~G.~Smith, Nature {\bf 298}, 825 (1982); J.~M.~Cordes,
  Astrophys. J. {\bf 311}, 183 (1986). 
%
%
B.~M.~S.~Hansen and E.~S.~Phinney,  
  Mon. Not. R. Astron. Soc. {\bf 291}, 569 (1997); 
J.~M.~Cordes and D.~F.~Chernoff,
  Astrophys. J. {\bf 505}, 315 (1998); 
  C.~Fryer, A.~Burrows, and W.~Benz,  Astrophys. J. {\bf 496}, 333 (1998).
  
\bibitem{astro_15} 
Z.~Arzoumanian, D.~F.~Chernoff and J.~M.~Cordes,
Astrophys. J. {\bf 568}, 289 (2002).

\bibitem{osc_matter_B}    V.~B.~Semikoz,
  Yad.\ Fiz.\  {\bf 46}, 1592 (1987); 
  J.~C.~D'Olivo, J.~F.~Nieves and P.~B.~Pal,
  Phys.\ Rev.\ D {\bf 40}, 3679 (1989);
  J.~C.~D'Olivo and J.~F.~Nieves,
  Phys.\ Rev.\ D {\bf 56}, 5898 (1997).



\bibitem{drt}
O.~F.~Dorofeev, V.~N.~Rodionov and I.~M.~Ternov, Sov. Astron. Lett. {\bf
11}, 123 (1985).


\bibitem{eq} A.~Vilenkin, Astrophys. J. {\bf 451}, 700 (1995); 
A.~Kusenko, G.~Segr\`e, and A.~Vilenkin, Phys. Lett. B 437, 359 (1998); 
P.~Arras and D.~Lai, astro-ph/9806285.

\bibitem{barkovich} 
M.~Barkovich, J.~C.~D'Olivo, R.~Montemayor and J.~F.~Zanella,
Phys.\ Rev.\ D {\bf 66}, 123005 (2002). 
%
%
M.~Barkovich, J.~C.~D'Olivo and R.~Montemayor,
Phys.\ Rev.\ D {\bf 70}, 043005 (2004)

\bibitem{Fryer}
  C.~L.~Fryer and A.~Kusenko,
  Astrophys.\ J.\ Suppl.\  {\bf 163}, 335 (2006).

\end{thebibliography}
\end{document}